# An Anonymous Authentication and Communication Protocol for Wireless Mesh Networks

Jaydip Sen

Innovation Lab, Tata Consultancy Services Ltd.,
Bengal Intelligent Park, Salt Lake Electronics Complex, Kolkata – 700091, India
`Jaydip.Sen@tcs.com`

**Abstract.** Wireless mesh networks (WMNs) have emerged as a key technology for next generation wireless broadband networks showing rapid progress and inspiring numerous compelling applications. A WMN comprises of a set of mesh routers (MRs) and mesh clients (MCs), where MRs are connected to the Internet backbone through the Internet gateways (IGWs). The MCs are wireless devices and communicate among themselves over possibly multi-hop paths with or without the involvement of MRs. User privacy and security have been primary concerns in WMNs due to their peer-to-peer network topology, shared wireless medium, stringent resource constraints, and highly dynamic environment. Moreover, to support real-time applications, WMNs must also be equipped with robust, reliable and efficient communication protocols so as to minimize the end–to-end latency and packet drops. Design of a secure and efficient communication protocol for WMNs, therefore, is of paramount importance. In this paper, we propose a security and privacy protocol that provides security and user anonymity while maintaining communication efficiency in a WMN. The security protocol ensures secure authentication and encryption in access and the backbone networks. The user anonymity, authentication and data privacy is achieved by application of a protocol that is based on Rivest's ring signature scheme. Simulation results demonstrate that while the protocols have minimal storage and communication overhead, they are robust and provide high level of security and privacy to the users of the network services.

## 1 Introduction

Wireless mesh networking has emerged as a promising concept to meet the challenges in next-generation wireless networks such as providing flexible, adaptive, and reconfigurable architecture while offering cost-effective solutions to service providers. WMNs are multi-hop wireless networks formed by mesh routers (which form a wireless mesh backbone) and mesh clients. The mesh routers provide a rich radio mesh connectivity which significantly reduces the up-front deployment cost of the network. Mesh routers are typically stationary and do not have power constraints. However, the clients are mobile and energy-constrained. Some mesh routers are designated as gate-way routers which are connected to the Internet through a wired backbone. A gateway router provides access to conventional clients and interconnects ad hoc, sensor, cellular, and other networks to the Internet. A mesh network can provide multi-hop communication paths between wireless clients, thereby serving as a community network, or can provide multi-hop paths between the client and the gateway router, thereby providing broadband Internet access to the clients.

As WMNs become an increasingly popular replacement technology for last-mile connectivity to the home networking, community and neighborhood networking, it is imperative to design an efficient resource management protocols for these networks. However, several vulnerabilities currently exist in various protocols for WMNs. These vulnerabilities can be exploited by the attackers to degrade the performance of a network. The absence of a central point of administration makes the WMN protocols vulnerable to various types of attacks. Security is therefore an issue which is of prime importance in WMNs [1]. Since in a WMN, traffic of an end user is relayed via multiple wireless mesh routers, preserving privacy of the user data is also a critical requirement [2]. Majority of the current security and privacy protocols for WMNs are extensions of protocols originally designed for mobile ad hoc networks (MANETs) and therefore their performances are suboptimal.

Keeping this problem in mind, this paper presents a novel security protocol for node authentication and message confidentiality for WMNs. In addition it also presents a user anonymization scheme that ensures secure authentication of the mesh clients (i.e., the user devices) while protecting their privacy.

The key contributions of the paper are as follows: (i) It proposes a novel security protocol for the mesh client nodes and the mesh routers. (ii) For protecting user privacy while providing a secure authentication framework for the mesh clients (user devices), it presents a novel anonymization scheme that utilizes the essential idea of Rivest group signature scheme [3].

The rest of this paper is organized as follows. Section 2 describes related work on routing in WMNs. Section 3 presents the details of the architecture of a WMN and the assumptions made for the development of the proposed protocols. Section 4 and Section 5 describe the proposed security and the privacy protocols respectively. Section 6 presents some performance results of the proposed scheme, and Section 7 highlights some future scope of work and concludes the paper.

## 2 Related Work

Since security and privacy are two extremely important issues in any communication network, researchers have worked on these two areas extensively. However, as comapred to MANETs and wireless sensor networks (WSNs), WMNs have received very little attention in this regard. . This section briefly discusses some of the existing mechanisms for ensuring security and privacy in communications in WMNs.

In [4], a standard mechanism has been proposed for client authentication and access control to guarantee a high-level of flexibility and transparency to all users in a wireless network. The users can access the mesh network without requiring any change in their devices and softwares. However, client mobility can pose severe prob-lems to the security architecture, especially when real-time traffic is transmitted. To cope with this problem, proactive key distribution has been proposed [5, 6].

Providing security in the backbone network for WMNs is another important challenge. Mesh networks typically employ resource constrained mobile clients, which are difficult to protect against removal, tampering, or replication. If the device can be remotely managed, a distant hacking into the device would work perfectly [7]. Accordingly, several research works have been done to investigate the use of cryptographic techniques to achieve secure communication in WMNs. In [8], a security architecture has been proposed that is suitable for multi-hop WMNs employing PANA (Protocol for carrying Authentication for Network Access) [9]. In the scheme, the wireless clients are authenticated on production of the cryptographic credentials necessary to create an encrypted tunnel with the remote access router to which they are associated. Even though such framework protects the confidentiality of the information exchanged, it cannot prevent adversaries to perform active attacks against the network itself. For instance, a malicious adversary can replicate, modify and forge the topology information exchanged among mesh devices, in order to launch a denial of service attack. Moreover, PANA necessitates the existence of IP addresses in all the mesh nodes, which is poses a serious constraint on deployment of this protocol.

Authenticating transmitted data packets is an approach for preventing unauthorized nodes to access the resources of a WMN. A light-weight hop-by-hop access protocol (LHAP) has been proposed for authenticating mobile clients in wireless dynamic environments, preventing resource consumption attacks [10]. LHAP implements light-weight hop-by-hop authentication, where intermediate nodes authenticate all the packets they receive before forwarding them. LHAP employs a packet authentication technique based on the use of one-way hash chains. Moreover, LHAP uses TESLA [11] protocol to reduce the number of public key operations for bootstrapping and maintaining trust between nodes.

In [12], a lightweight authentication, authorization and accounting (AAA) infrastructure is proposed for providing continuous, on-demand, end-to-end security in heterogeneous networks including WMNs. The notion of a security manager is used through employing an AAA broker. The broker acts as a settlement agent, providing security and a central point of contact for many service providers.

The issue of user privacy in WMNs has also attracted the attention of the research community. In [2], a light-weight privacy preserving solution is presented to achieve well-maintained balance between network performance and traffic privacy preservations. At the center of the solution is of information-theoretic metric called traffic entropy, which quantifies the amount of information required to describe the traffic pattern and to characterize the performance of traffic privacy preservation. The authors have also presented a penalty-based shortest path routing algorithm that maximally preserves traffic privacy by minimizing the mutual information of traffic entropy observed at each individual relaying node, meanwhile controlling performance degradation within the acceptable region. Extensive simulation study proves the soundness of the solution and its resilience to cases when two malicious observers collude. However, one of the major problems of the solution is that the algorithm is evaluated in a single-radio, single channel WMN. Performance of the algorithm in multiple radios, multiple channels –scenario will be a really questionable issue. Moreover, the solution has a scalability problem.

In [13], a mechanism is proposed with the objective of hiding an active node that connects to a gateway router, where the active mesh node has to be anonymous. A novel communication protocol is designed to protect the node's privacy using both cryptography and redundancy. This protocol uses the concept of onion routing [14]. A mobile user who requires anonymous communication sends a request to an onion router (OR). The OR acts as a proxy to the mobile user and constructs a onion route consisting of other ORs using the public keys of the routers. The onion is constructed such that the inner most part is the message for the intended destination, and the message is wrapped by being encrypted using the public keys of the ORs in the route. The mechanism protects the routing information from insider and outsider attack. However, it has a high computation and communication overhead.

None of the above propositions, however, addresses all the security problems of a typical WMN. Most of the schemes handle security issues at a specific layer, and therefore, fail to provide a multi-layer attack on the protocol stack of a WMN. This paper proposes a security and privacy framework that addresses issues both at the access and the backbone networks while not affecting the network performance.

## 3  WMN Security Architecture

In this section, we first present a standard architecture of a typical WMNS for which we propose a security and privacy protocol. The architecture is a very generic one that represents majority of the real-world deployment scenarios for WMNs. The architecture of a hierarchical WMN consists of three layers as shown in Fig. 1. At the top layers are the *Internet gateways* (IGWs) that are connected to the wired Internet. They form the backbone infrastructure for providing Internet connectivity to the elements in the second level. The entities at the second level are called wireless *mesh routers* (MRs) that eliminate the need for wired infrastructure at every MR and forward their traffic in a multi-hop fashion towards the IGW. At the lowest level are the *mesh clients* (MCs) which are the wireless devices of the users. Internet connectivity and peer-to-peer communications inside the mesh are two important applications for a WMN. Therefore design of an efficient and low-overhead communication protocol which ensure security and privacy of the users is a critical requirement which poses significant research challenges. For design of the proposed protocol and to specify the WMN scenario, the following assumptions are made.

(1) Each MR which is authorized to join the wireless backbone (through the IGWs), has two certificates to prove its identity. One certificate is used during the authentication phase that occurs when a new node joins the network. EAP-TLS [15] for 802.1X authentication is used for this purpose since it is the strongest authentication method provided by EAP [15], whereas the second certificate is used for the authentication with the *authentication server* (AS).

(2) The certificates used for authentication with the RADIUS server and the AS are signed by the same *certificate authority* (CA). Only recognized MRs are authorized to join the backbone.

(3) Synchronization of all MRs is achieved by use of the *network time protocol* (NTP) protocol [16].

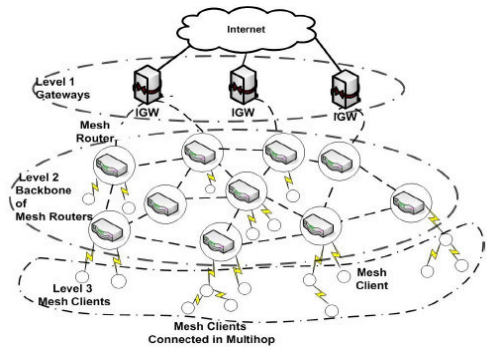

**Fig. 1**. The three-tier architecture of a wireless mesh network (WMN)

The proposed security protocol serves the dual purpose of providing security in the access network (i.e., between the MCs and the MRs) and the backbone network (i.e., between the MRs and the IGWs). These are described the following sub-sections.

### 3.1  Access Network Security

The access mechanism to the WMN is assumed to be the same as that of a *local area network* (LAN), where mobile devices authenticate themselves and connect to an *access point* (AP). This allows the users to the access the services of the WMN exploiting the authentication and authorization mechanisms without installing any additional software. It is evident that such security solution provides protection to the wireless links between the MCs and the MRs. A separate security infrastructure is needed for the links in the backbone networks. This is discussed in Section 3.2.

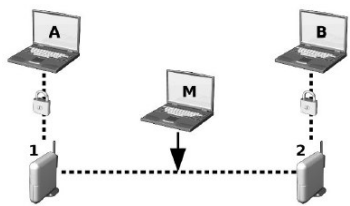

**Fig. 2**. Secure information exchange among the MCs *A* and *B* through the MRs 1 and 2

Fig. 2 illustrates a scenario where users *A* and *B* are communicating in a secure way to MRs 1 and 2 respectively. If the wireless links are not protected, an intruder *M* will be able to eavesdrop on and possibly manipulate the information being exchanged over the network. This situation is prevented in the proposed security scheme which encrypts all the traffic transmitted on the wireless link using a stream cipher in the data link layer of the protocol stack.

## 3.2 Backbone Network Security

For providing security for the traffic in the backbone network, a two-step approach is adopted. When a new MR joins the network, it first presents itself as an MC and completes the association formalities. It subsequently upgrades its association by succsssfully authenticating to the AS. In order to make such authentication process efficient in a high mobility scenario, the key management and distribution processes have been designed in a way so as to minimize the effect of the authentication overhead on the network performance. The overview of the protocol is discussed as follows.

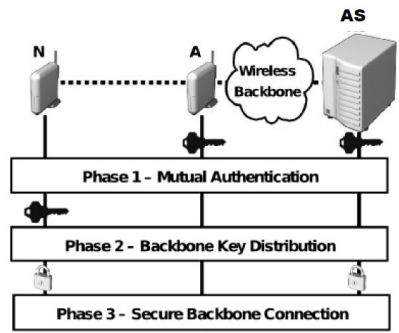

**Fig. 3**. Steps performed by a new MR (*N*) using backbone encrypted traffic to join the WMN

Fig. 3 shows the three phases of the authentication process that a MR (say *N*) undergoes. When *N* wants to join the network, it scans all the radio channels to detect any MR that is already connected to the wireless backbone. Once such an MR (say *A*) is detected, *N* requests *A* for access to network services including authentication and key distribution. After connecting to *A*, *N* can perform the tasks prescribed in the IEEE 802.11i protocol to complete a mutual authentication with the network and establish a security association with the entity to which it is physically connected. This completes the Phase I of the authentication process. Essentially, during this phase, a new MR performs all the steps that an MC has to perform to establish a secure channel with an MR for authentication and secure communication over the WMN.

During Phase II of the authentication process, the MRs use the TLS protocol. Only authorized MRs that have the requisite credentials can authenticate to the AS and obtain the cryptographic credentials needed to derive the key sequence used to protect the wireless backbone. In the proposed protocol, an end-to-end secure channel between the AS and the MR is established at the end of a successful authentication through which the cryptographic credentials can be exchanged in a secure way.

To eliminate any possibility of the same key being used over a long time, two protocols are proposed for secure key management. These protocols are presented in Section 4. As mentioned earlier in this section, all the MRs are assumed to be synchronized with a central server using the NTP protocol.

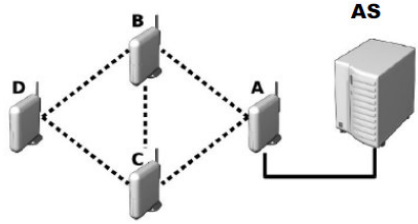

**Fig. 4.** Autonomous configuration of the MRs in the proposed security scheme

Fig. 4 shows a collection of four MRs connected with each other by five wireless links. The MR *A* is connected with the AS by a wired link. At the time of network bootstrapping, only node *A* can connect to the network as an MR, since it is the only node that can successfully authenticate to the AS. Nodes *B* and *C* which are neighbors of *A* then detect a wireless network to which can connect and perform the authentication process following the IEEE 802.11i protocol. At this point of time, nodes *B* and *C* are successfully authenticated as MCs. After their authentication as MCs, nodes *B* and *C* are allowed to authenticate to the AS and request the information used by *A* to produce the currently used cryptographic key for communication in the network. After having derived such key, both *B* and *C* will be able to communicate with each other, as well as with node *A*, using the ad hoc mode of communication in the WMN. At this stage, *B* and *C* both have full MR functionalities. They will be able to turn on their access interface for providing node *D* a connection to the AS for joining the network.

## 4 The Key Distribution Protocol

In this section, the details of the proposed key distribution and management protocol are presented. The protocol is essentially a server-initiated protocol [17] and provides the clients (MRs and MCs) flexibility and autonomy during the key generation.

### 4.1 Server Initiated Key Management Protocol

In the proposed key management protocol delivers the keys to all the MRs from the AS in a reactive manner. The keys are used subsequently by the MRs for a specific time interval in their message communications to ensure integrity and confidentiality of the messages. After the expiry of the time interval for validity of the keys, the existing keys are revoked and new keys are generated by the AS. Fig. 5 depicts the message exchanges between the MRs and the AS during the execution of the protocol.

A newly joined MR, after its successful mutual authentication with a central server, sends its first request for key list (and its time of generation) currently being used by other existing MRs in the wireless backbone. Let us denote the *key list timestamp* as $TS_{KL}$. Let us define a *session* as the maximum time interval for validity of the key list currently being used by each node MR and MC). We also define the duration of a session as the product of the *cardinality of the key list* (i.e., the number of the keys in the key list) and the longest time interval of validity of a key (the parameter *timeout* in

Fig. 5). The validity of a key list is computed from the time instance when the list is generated (i.e., $TS_{KL}$) by the AS. An MR, based on the time instance at which it joins the backbone ($t_{now}$ in Fig. 5), can find out the key (from the current list) being used by its peers ($key_{idx}$) and the interval of validity of the key ($T_i$) using (1) and (2) as follows:

$$key_{idx} = \left\lfloor \frac{t_{now} - TS_{KL}}{timeout} \right\rfloor + 1 \qquad (1)$$

$$T_i = key_{idx} * timeout - (t_{now} - TS_{KL}) \qquad (2)$$

In the proposed protocol, each WMN node requests the AS for the *key list* that will be used in the next session before the expiry of the current session. This is feature is essential for nodes which are located multiple hops away from the AS, since, responses from the AS take longer time to reach these nodes. The responses may also get delayed due to fading or congestion in the wireless links. If the nodes send their requests for key list to the AS just before expiry of the current session, then due to limited time in hand, only the nodes which have good quality links with the AS will receive the key list. Hence, the nodes which will fail to receive responses for the server will not be able to communicate in the next session due to non-availability of the current key list. This will lead to an undesirable situation of network partitioning.

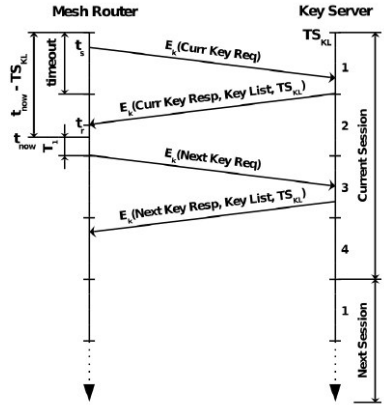

**Fig. 5**. The message exchanges between an MR and the AS in the key management protocol.

The *key index* value that triggers the request from the nodes to the server can be set equal to the difference between the *cardinality of the list* and a *correction factor*. The correction factor can be estimated based on parameters like the network load, the distance of the node from the AS and the time required for the previous response.

In the proposed protocol, the correction factor is estimated based on the time to receive the response from the AS using (3), where $t_s$ is the time instance when the first key request was sent, $t_r$ is the time instance when the key response was received from the AS, and *timeout* is the validity period of the key. Therefore, if a node fails to receive a response (i.e., the key list) from the AS during timeout, and takes a time $t_{last}$, it must send the next request to the AS before setting the last key.

$$c = \left\lceil \frac{t_{last} - timeout}{timeout} \right\rceil \quad \text{if } t_{last} \geq timeout$$
$$= 0 \quad \text{if } t_{last} < timeout$$

$$t_{last} = t_r - t_s \tag{3}$$

The first request of the key list sent by the new node to the AS is forwarded by the peer to which it is connected as an MC through the wireless access network. However, the subsequent requests are sent directly over the wireless backbone.

## 5  The Privacy and Anonymity Protocol

As mentioned in Section 1, to ensure privacy of the users, the proposed security protocol is complemented with a privacy protocol so as to ensure user anonymity and privacy. The same *authentication server* (AS) used in the security protocol is used for managing the key distribution for preserving the privacy. To enable user authentication and anonymity, a novel protocol has been designed extending the *ring signature authentication* scheme in [18]. It is assumed that a symmetric encryption algorithm $E$ exists such that for any key $k$, the function $E_k$ is a permutation over $b$-bit strings. We also assume the existence of a family of *keyed combining functions* $C_{k,v}(y_1, y_2, ...., y_n)$, and a publicly defined *collision-resistant* hash function $H(.)$ that maps arbitrary inputs to strings of constant length which are used as keys for $C_{k,v}(y_1, y_2, ...., y_n)$ [3]. Every keyed combining function $C_{k,v}(y_1, y_2, ...., y_n)$ takes as input the key $k$, an initialization $b$-bit value $v$, and arbitrary values $y_1, y_2, ...., y_n$. A user $U_i$ who wants to generate a session key with the authentication server, uses a ring of $n$ logged-on-users and performs the following steps.

*Step 1*: $U_i$ chooses the following parameters: (i) a large prime $p_i$ such that it is hard to compute discrete logarithms in $GF(p_i)$, (ii) another large prime $q_i$ such that $q_i \mid p_i - 1$, and (iii) a generator $g_i$ in $GF(p_i)$ with order $q_i$.

*Step 2*: $U_i$ chooses $x_{A_i} \in Z_{q_i}$ as his private key, and computes the public key $y_{A_i} = g_i^{x_{Ai}} \mod p_i$.

*Step 3*: $U_i$ defines a trap-door function $f_i(\alpha, \beta) = \alpha . y_{Ai}^{\alpha \mod q_i} . g_i^{\beta} \mod p_i$. Its inverse function $f_i^{-1}(y)$ is defined as $f_i^{-1}(y) = (\alpha, \beta)$, where $\alpha$ and $\beta$ are computed as follows ($K$ is a random integer in $Z_{q_i}$.

$$\alpha = y_{Ai} . g_i^{-K.(g_i^K \mod p_i) \mod q_i} \mod p_i \tag{4}$$

$$\alpha^* = \alpha \bmod q_i \tag{5}$$

$$\beta = K \cdot (g_i^K \bmod p_i) - x_{A_i} \cdot \alpha^* \bmod q_i \tag{6}$$

$U_i$ makes $p_i$, $q_i$, $g_i$ and $y_{A_i}$ public, and keeps $x_{A_i}$ as secret.

The *authentication server* (*AS*) chooses: (i) a large prime $p$ such that it is hard to compute discrete logarithms in $GF(p)$, (ii) another large prime $q$ such that $q \mid p - 1$, (iii) a generator $g$ in $GF(p)$ with order $q$, (iv) a random integer $x_B$ from $Z_q$ as its private key. *AS* computes its public key $y_B = g^{x_B} \bmod p$ and publishes ($y_B$, $p$, $q$, $g$).

**Anonymous authenticated key exchange**: The key-exchange is initiated by the user $U_i$ and involves three rounds to compute a secret session key between $U_i$ and *AS*. The operations in these three rounds are as follows:

*Round 1*: When $U_i$ wants to generate a session key on the behalf of $n$ ring users $U_1$, $U_2$, .....$U_n$, where $1 \le i \le n$, $U_i$ does the following:

(i) $U_i$ chooses two random integers $x_1, x_A \in Z_q^*$ and computes the following: $R = g^{x_1} \bmod p$, $Q = y_B^{x_1} \bmod p \bmod q$, $X = g^{x_a} \bmod p$ and $l = H(X, Q, V, y_B, I)$.

(ii) $U_i$ Chooses a pair of values $(\alpha_t, \beta_t)$ for every other ring member $U_t$ $(1 \le t \le n, t \ne k)$ in a pseudorandom way, and computes $y_t = f_t(\alpha_t, \beta_t) \bmod p_t$.

(iii) $U_i$ randomly chooses a $b$-bit initialization value $v$, and finds the value of $y_i$ from the equation $C_{k,v}(y_1, y_2, \ldots y_n) = v$.

(iv) $U_i$ computes $(\alpha_i, \beta_i) = f_i^{-1}(y_i)$ by using the trap-door information of $f_i$. First, it chooses a random integer $K \in Z_{q_i}$, computes $\alpha_i$ using (6), and keeps $K$ secret. It then computes $\alpha_i^*$ using (5) and finally computes $\beta_i$ using (6).

(v) $(U_1, U_2, .., U_n, v, V, R, (\alpha_1, \beta_1), (\alpha_2, \beta_2), .., (\alpha_n, \beta_n))$ is the ring signature $\sigma$ on $X$.

Finally, $U_i$ sends $\sigma$ and $I$ to the server *AS*.

*Round 2*: *AS* does the following to recover and verify $X$ from the signature $\sigma$.

(i) *AS* computes $Q = R^{x_B} \bmod p \bmod q$, recovers $X$ using $X = V \cdot g^Q \bmod p$ and hashes $X$, $Q$, $V$ and $y_b$ to recover $l$, where $l = H(X, Q, V, y_B, I)$.

(ii) *AS* computes $y_t = f_i(\alpha_t, \beta_t) \bmod p_i$, for $t = 1, 2, \ldots n$.

(iii) *AS* checks whether $C_{k,v}(y_1, y_2, \ldots y_n) = v$. If it is true, *AS* accepts $X$ as valid; otherwise, *AS* rejects $X$. If $X$ is valid, *AS* chooses a random integer $x_b$ from $Z_q^*$, and computes the following: $Y = g^{x_b} \bmod p$  $K_s = X^{x_b} \bmod p$ and $h = H(K_s, X, Y, I')$. *AS* sends $\{h, Y, I'\}$ to $U_i$.

*Round 3*: $U_i$ verifies whether $K_s'$ is from the server *AS*. For this purpose, $U_i$ computes $K_s' = Y^{x_a} \bmod p$, hashes $K$, $X$, $Y$ to get $h'$ using $h' = H(K_s', X, Y, I')$. If $h' ?= h$, $U_i$ accepts $K_s$ as the session key.

*Security analysis*: The key exchange scheme satisfies the following requirements.

*User anonymity*: For a given signature *X*, the server can only be convinced that the ring signature is actually produced by at least one of the possible users. If the actual user does not reveal the seed *K*, the server cannot determine the identity of the user. The strength of the anonymity depends on the security of the pseudorandom number generator. It is not possible to determine the identity of the actual user in a ring of size *n* with a probability greater than 1/*n*. Since the values of *k* and *v* are fixed in a ring signature, there are $(2^b)^{n-1}$ number of $(x_1, x_2, ... x_n)$ that satisfy the equation $C_{k,v}(y_1, y_2, ... y_n) = v$, and the probability of generation of each $(x_1, x_2, ... x_n)$ is the same. Therefore, the signature can't leak the identity information of the user.

*Mutual authentication*: In the proposed scheme, not only the server verifies the users, but the users can also verify the server. Because of the hardness of inverting the hash function *f(.)*, it is computationally infeasible for the attacker to determine $(\alpha_i, \beta_i)$, and hence it is infeasible for him to forge a signature. If the attacker wants to masquerade as the *AS*, he needs to compute $h = H(K_s, X, Y)$. He requires $x_B$ in order to compute *X*. However, $x_B$ is the private key of *AS* to which the attacker has no access.

*Forward secrecy*: The forward secrecy of a scheme refers to its ability to defend leaking of its keys of previous sessions when an attacker is able to catch hold of the key of a particular session. The forward secrecy of a scheme enables it to prevent *replay attacks*. In the proposed scheme, since $x_a$ and $x_b$ are both selected randomly, the session key of each period has not relation to the other periods. Therefore, if the session key generated in the period *j* is leaked, the attacker can not get any information of the session keys generated before the period *j*. The proposed protocol is, therefore, resistant to replay attack.

## 6 Performance Evaluation

The proposed security and privacy protocols have been implemented in the Qualnet network simulator, version 4.5 [19]. The simulated network consists of 50 nodes randomly distributed in the simulation area forming a dense WMN. The WMN topology is shown in Fig. 6, in which 5 are MRs and remaining 45 are MCs. Each MR has 9 MCs associated with it. To evaluate the performance of the security protocol, first the network is set as a full-mesh topology, where each MR (and also MC) is directly connected to two of its neighbors. In such as scenario, the throughput of a TCP connection established over a wireless link is measured with the security protocol activated in the nodes. The obtained results are then compared with the throughput obtained on the same wireless link protected by a static key to encrypt the traffic.

After having 10 simulation runs, the average throughput of a wireless link between a pair of MRs was found to be equal to 30.6 MBPS, when the link is protected by a static key. However, the average throughput for the same link was 28.4 MBPS when the link was protected by the proposed security protocol. The results confirm that the protocol does not cause any significant overhead on the performance of the wireless link, since the throughput in a link on average decreased by only 7%.

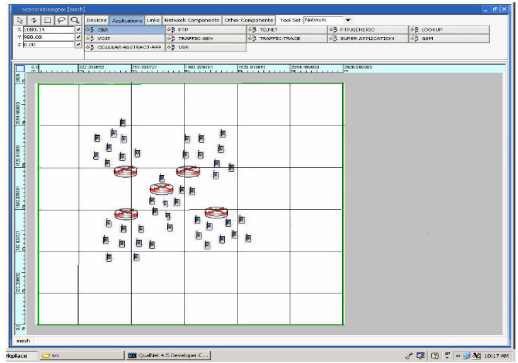

**Fig. 6.** The simulated network topology in Qualnet Simulator

The impact of the security protocol for key generation and revocation on packet drop rate in real-time applications is also studied in the simulation. For this purpose, a VoIP application is invoked between two MRs which generated UDP traffic in the wireless link. The packet drop rates in wireless link when the link is protected with the proposed security protocol and when the link is protected with a static key. The transmission rate was set to 1 MBPS. The average packet drop rate in 10 simulation runs was found to be only 4%. The results clearly demonstrate that the proposed security scheme has no adverse impact on packet drop rate even if several key switching (regeneration and revocation) operations are carried out.

The performance of the privacy protocol is also analyzed in terms of its storage, communication overhead. Both storage and communication overhead were found to increase linearly with the number of nodes in the network. In fact, it has been analytically shown that overhead due to cryptographic operation on each message is: $60n + 60$ bytes, where $n$ represents the number of public key pairs used to generate the ring signature [20]. It is clear that the privacy protocol has a low overhead.

## 7 Conclusion and Future Work

WMNs have become an important focus area of research in recent years owing to their great promise in realizing numerous next-generation wireless services. Driven by the demand for rich and high-speed content access, recent research has focused on developing high performance communication protocols, while security and privacy issues have received relatively little attention. However, given the wireless and multi-hop nature of communication, WMNs are subject to a wide range of security and privacy threats. This paper has presented a security and user-privacy preserving protocol for WMNs. The proposed security protocol ensures security in both the access and the backbone networks, whereas the privacy protocol enables anonymous authentication of the users. Simulation results have shown the effectiveness of the protocol. Future research issues include the study of a distributed and collaborative system where the authentication service is provided by a dynamically selected set of MRs. The integration with the current centralized scheme would increase the robustness of the proposed protocol, maintaining a low overhead since MRs would use the distributed service only when the central server is not available.